\documentclass[aps,showpacs,amsmath,amssymb,12pt]{revtex4}
\usepackage{latexsym}
\usepackage{bm}

\def\HollowBox #1#2{{\dimen0=#1 \advance\dimen0 by -#2
       \dimen1=#1 \advance\dimen1 by #2
        \vrule height #1 depth #2 width #2
        \vrule height 0pt depth #2 width #1
        \llap{\vrule height #1 depth -\dimen0 width \dimen1} 
       \hskip -#2
       \vrule height #1 depth #2 width #2}}
\def\BOX{\HollowBox{.100in}{.010in}}

\begin{document}

\title{Mass and angular momentum of asymptotically AdS or flat 
solutions in the topologically massive gravity}

\author{Serkay {\" O}lmez}
\email{olmezs@newton.physics.metu.edu.tr}
\affiliation{Department of Physics, Faculty of Arts and  Sciences,\\
             Middle East Technical University, 06531, Ankara, Turkey}

\author{{\" O}zg{\" u}r Sar{\i}o\u{g}lu}  
\email{sarioglu@metu.edu.tr}
\affiliation{Department of Physics, Faculty of Arts and  Sciences,\\
             Middle East Technical University, 06531, Ankara, Turkey}

\author{Bayram Tekin}  
\email{btekin@metu.edu.tr}
\affiliation{Department of Physics, Faculty of Arts and  Sciences,\\
             Middle East Technical University, 06531, Ankara, Turkey}

\date{\today}

\begin{abstract}
We study the conserved charges of supersymmetric solutions in the
topologically massive gravity theory for both asymptotically flat
and constant curvature geometries.
\end{abstract}


\maketitle

\section{\label{intro} Introduction}

In the presence of a cosmological constant, the source-free field 
equations of the (2+1 dimensional) topologically massive gravity 
(TMG) theory read
\[ R_{\mu\nu} - \frac{1}{2} \, g_{\mu\nu} \, R + \Lambda \, g_{\mu\nu} 
+ \frac{1}{\mu} \, C_{\mu\nu} = 0 \, , \qquad 
C^{\mu\nu} \equiv \frac{1}{\sqrt{-g}} \, \epsilon^{\mu\alpha\beta} \, 
\nabla_{\alpha} \left( R^{\nu}\,_{\beta} - \frac{1}{4} \, 
\delta^{\nu}\,_{\beta} \, R \right) \, , \]
where the Cotton tensor $C_{\mu\nu}$ is the three dimensional analogue of the
Weyl tensor and is symmetric, traceless and identically conserved; the 
parameter $\mu$ is the coupling constant for the gravitational Chern-Simons
term in the action and corresponds to the mass of the linearized TMG 
excitations at $\Lambda=0$ (see \cite{des} for details and for 
$\Lambda \neq 0$ see \cite{dba1}). A minimally supersymmetric extension 
of this theory was also constructed long time ago \cite{deskay}.

The well known BTZ metric \cite{btz}, as well as the Anti-de Sitter (AdS) 
and the Schwarzschild-dS spacetimes, are solutions to TMG theory with a
cosmological constant in a `trivial' manner since their Cotton tensors vanish
identically. There are, however, other known `nontrivial' solutions; {\it i.e.}
those spacetimes that obey the `full' TMG equations, not just their Einstein 
part (or the cosmological Einstein part in the relevant cases) alone. The 
first example that we know is Deser's gravitational anyons which are only
solutions of the linearized TMG equations \cite{any}. The `fully' nonlinear,
`nontrivial' solutions include the Vuorio solution \cite{vuo} and its 
generalization to solutions with a constant twist \cite{per}. There are 
also exact static/stationary solutions for spinning point sources for which 
the spin and the mass of the sources obey a certain relation \cite{ort}, 
\cite{par}. A class of cosmological-type solutions is given by finite action 
exact solutions of TMG \cite{nub} that are also useful for a classification 
of homogeneous solutions. There also exists a two parameter solution to 
TMG theory with a cosmological constant which seems to have properties 
similar to the BTZ solution \cite{nut}, \cite{gur}. This solution is not 
asymptotically AdS in its original form, but then its asymptotically AdS 
form, which is obtained by imposing a certain relation between $\Lambda$ 
and $\mu$, is equivalent to the BTZ metric. Another class of solutions 
which asymptotically approach extremal BTZ black holes, but are geodesically 
complete with no event horizons, were also given \cite{cle}. Finally, the 
first nontrivial example of a solution to TMG that preserves half of 
supersymmetry was found in \cite{ds}; moreover, the solutions in 
\cite{cle} seem to be related to the supersymmetric solutions by a 
certain choice of parameters and a coordinate transformation. Another 
two-parameter family of black hole solutions that are obtained by an
analytical continuation of the Vuorio solution, but fail to be asymptotically
AdS like their ancestor, was recently given in \cite{dal}.

In a recent work \cite{db1}, a concrete and rigorous definition of
conserved gravitational charges (particularly energy and angular momentum)
were given in a `surface' integral form about their flat or asymptotically
AdS backgrounds in TMG theory. It is only natural to consider the exact 
solutions listed in the previous paragraph as explicit examples whose
gravitational charges can be calculated a la \cite{db1}. We do this for 
the BTZ and the only nontrivial supersymmetric solutions of TMG that are
asymptotically AdS or flat in this paper. This should also help to better 
understand the physical properties of these examples and to clarify the 
physical meanings of some of the parameters that explicitly show up in them. 

\section{\label{conchar} The conserved gravitational charges of the TMG theory}

Let us start by giving a brief outline of how gravitational charges are
defined in TMG. (We refer the reader to \cite{db1} and \cite{db2} for 
details.) Assume that the deviation, $h_{\mu\nu}$, of the actual spacetime
metric \( g_{\mu\nu} = \bar{g}_{\mu\nu} + h_{\mu\nu} \) from an 
asymptotically AdS metric (or the background) $\bar{g}_{\mu\nu}$, which
obeys
\[ \bar{R}_{\mu\alpha\nu\beta} = \Lambda \, 
(\bar{g}_{\mu\nu} \, \bar{g}_{\alpha\beta} - 
\bar{g}_{\mu\beta} \, \bar{g}_{\alpha\nu}) \, , \quad
\bar{R}_{\mu\nu} = 2 \Lambda \, \bar{g}_{\mu\nu} \, , \quad
\bar{R} = 6 \Lambda \, , \]
is employed for constructing ``linearized gravity'' in the usual sense with
the usual assumptions \cite{db2}. Then the `linearized' part of the Ricci
tensor \footnote{Here \( h \equiv h_{\mu\nu} \, \bar{g}^{\mu\nu} \),
all indices are raised and lowered with the background metric
$\bar{g}_{\mu\nu}$ and also all covariant differentiations are carried 
with respect to $\bar{g}_{\mu\nu}$.}
\[ R_{\mu\nu}^{L} = \frac{1}{2} (- \bar{\BOX} \, {h}_{\mu\nu} 
- \bar{\nabla}_{\mu} \, \bar{\nabla}_{\nu} \, h + \bar{\nabla}^{\sigma} \,
\bar{\nabla}_{\nu} \, h_{\sigma\mu} + \bar{\nabla}^{\sigma} \,
\bar{\nabla}_{\mu} \, h_{\sigma\nu}) \, , \]
and the linearized Ricci scalar
\[ R^{L} \equiv (R_{\mu\nu} \, g^{\mu\nu})^{L} = 
R_{\mu\nu}^{L} \, \bar{g}^{\mu\nu} - 2 \Lambda \, h = - \bar{\BOX} \, h
+ \bar{\nabla}_{\mu} \, \bar{\nabla}_{\nu} \, \bar{h}^{\mu\nu} 
- 2 \Lambda \, h \, , \]
can be used for finding the linearized cosmological Einstein and the
Cotton tensors as
\begin{eqnarray*}
{\cal G}_{\mu\nu} & \equiv & (G_{\mu\nu} + \Lambda \, g_{\mu\nu})^{L} =
R_{\mu\nu}^{L} - \frac{1}{2} \, \bar{g}_{\mu\nu} \,
R^{L} - 2 \Lambda \, {h}_{\mu\nu} \, , \\
C^{\mu\nu}_{L} & = & \frac{1}{\sqrt{-\bar{g}}} \, \epsilon^{\mu\alpha\beta} \,
\bar{g}_{\beta\sigma} \, \bar{\nabla}_{\alpha} \, \left( R^{\sigma\nu}_{L}
- 2 \, \Lambda \, h^{\sigma\nu} - \frac{1}{4} \, \bar{g}^{\sigma\nu} 
\, R_{L} \right) \, . 
\end{eqnarray*}
Now one can find a background conserved and gauge invariant charge
(corresponding to each background Killing vector $\bar{\xi}^{\mu}$) which is
given as the sum of the following three terms:
\begin{equation}
Q^{\mu} (\bar{\xi}) = \frac{1}{8 \pi G} \, \oint_{\partial {\cal M}} \,
dS_{i} \, \left( q^{\mu i}_{E} (\bar{\xi}) + \frac{1}{2 \mu} \, 
q^{\mu i}_{E} (\bar{\Xi}) + \frac{1}{2 \mu} \, q^{\mu i}_{C} (\bar{\xi})
\right) \, , \label{charge}
\end{equation}
where \footnote{Here one in fact has 
\( {\cal G}^{\mu\nu} \equiv (G^{\mu\nu} + \Lambda \, g^{\mu\nu})^{L} \) but
then since \( \bar{G}_{\mu\nu} + \Lambda \, \bar{g}_{\mu\nu} = 0 \), 
moving the indices of the linearized cosmological Einstein tensor ${\cal G}$ 
can equivalently be carried out with the background metric $\bar{g}_{\mu\nu}$.}
\begin{eqnarray}
q^{\mu i}_{E} (\bar{\xi}) & \equiv & \sqrt{-\bar{g}} \left( 
\bar{\xi}_{\nu} \, \bar{\nabla}^{\mu} \, h^{i \nu} -
\bar{\xi}_{\nu} \, \bar{\nabla}^{i} \, h^{\mu\nu} +
\bar{\xi}^{\mu} \, \bar{\nabla}^{i} \, h -
\bar{\xi}^{i} \, \bar{\nabla}^{\mu} \, h \right.  \nonumber \\
& & \quad \qquad \left. + h^{\mu\nu} \, \bar{\nabla}^{i} \, \bar{\xi}_{\nu}
- h^{i \nu} \, \bar{\nabla}^{\mu} \, \bar{\xi}_{\nu}
+ \bar{\xi}^{i} \, \bar{\nabla}_{\nu} \, h^{\mu\nu}
- \bar{\xi}^{\mu} \, \bar{\nabla}_{\nu} \, h^{i \nu}
+ h \, \bar{\nabla}^{\mu} \, \bar{\xi}^{i} \right) \, , \label{einchar} \\
q^{\mu i}_{C} (\bar{\xi}) & \equiv & 
\epsilon^{\mu i \beta} \, {\cal G}_{\nu\beta} \, \bar{\xi}^{\nu}
+ \epsilon^{\nu i \beta} \, {\cal G}^{\mu}\,_{\beta} \, \bar{\xi}_{\nu}
+ \epsilon^{\mu\nu\beta} \, {\cal G}^{i}\,_{\beta} \, \bar{\xi}_{\nu} \, ,
\label{cotchar}
\end{eqnarray}
and 
\( \bar{\Xi}^{\beta} \equiv \epsilon^{\alpha\nu\beta} \, 
\bar{\nabla}_{\alpha} \, \bar{\xi}_{\nu} / \sqrt{-\bar{g}} \)
is another background Killing vector constructed out of $\bar{\xi}$. Here
${\cal M}$ is a spatial 2-dimensional hypersurface, $\partial {\cal M}$
is its 1-dimensional boundary and $i$ denotes the space direction orthogonal
to the boundary $\partial {\cal M}$ with the corresponding line element
$dS_{i}$. $G$ denotes the 3-dimensional Newton's constant and the charge
has been normalized by the overall factor $8 \pi G$ in (\ref{charge}).

\section{\label{btzbh} The BTZ black hole}

To set the stage properly, let us take the BTZ solution \cite{btz}
\begin{equation} 
ds^2 = \left( M- \frac{r^2}{\ell^2} \right) dt^2 - J \, dt \, d\phi
+ r^2 \, d\phi^2 + \frac{dr^2}{- M + \frac{r^2}{\ell^2} + \frac{J^2}{4r^2}}
\label{btzmet}
\end{equation}
as a first example. The correct black hole vacuum background is found by
setting $M=0$, $J=0$ in (\ref{btzmet}) (see \cite{btz} for a discussion on
this) which is clearly locally AdS:
\[ ds^2 = - \frac{r^2}{\ell^2} \, dt^2 + \frac{\ell^2}{r^2} \, dr^2 
+ r^2 \, d\phi^2 \, . \]
The timelike \( \bar{\xi}^{\mu} = (- \partial/\partial t)^{\mu} \) and
the spacelike \( \bar{\zeta}^{\mu} = (\partial/\partial \phi)^{\mu} \) 
Killing vectors can be used in finding the conserved energy and the 
angular momentum, respectively. The surface integral (\ref{charge}) at
some finite distance $r$ from the origin yields the following 
{\it non gauge-invariant} quantities, which afterwards give the `true' energy 
and angular momentum that are only to be measured at infinity 
\footnote{Throughout we have chosen the Newton constant $G$ in 
(\ref{charge}) such that one finds the usual ADM pair $(M,J)$ in the 
limit $\mu \rightarrow \infty$.}:
\begin{eqnarray*}
E(r) & = & \frac{4 r^4 (J- \mu M \ell^2) + J \ell^2 r^2 (\mu J - 4 M) + 
J^3 \ell^2}{-4 \mu \ell^2 r^4 + 4 \mu M \ell^4 r^2 - \mu J^2 \ell^4} \, , \\
L(r) & = & \frac{8 r^5 (\mu J - M) + J r^3 (J - 8 \mu M \ell^2) 
+ J^2 \ell^2 r (2 \mu J -M)}{2 \mu r (4 r^4 - 4 M \ell^2 r^2 + J^2 \ell^2)} 
\, .
\end{eqnarray*}
As a result, one obtains the energy and the angular momentum in the 
limit as $r \rightarrow \infty$ to be
\[ E = M - \frac{J}{\mu \ell^2} \qquad \mbox{and} \qquad 
L = J - \frac{M}{\mu} \, . \]

These quantities \cite{dal, dkt} are obviously different from the ADM charges 
of the BTZ black hole \cite{btz}; the Cotton part clearly has a nontrivial 
contribution to the conserved charges. Amusingly enough, the angular momentum 
vanishes when the two parameters $M$ and $J$ are related by $M=\mu J$, in 
which case \( E=M (1-1/(\mu^2 \ell^2)) \). Thus, if furthermore 
$\mu^2 \ell^2 =1$, then the BTZ black hole is left with no `energy' and
`angular momentum' in the TMG context!

As a brief remark on the charged version of the BTZ solution \cite{mtz},
we note that since the ``electric potential'' rises logarithmically in
$D=3$, even a cursory look suggests that a single charged black hole will
have divergent energy. In fact the authors of \cite{mtz} define the energy
of their charged rotating solution only upto an infinite constant factor.
In this respect, the gauge invariant energy in the sense of \cite{db1} is 
naturally found to be divergent for this case.

\section{\label{susytmg} The supersymmetric solution}

The half supersymmetry preserving solution given in \cite{ds} is 
described by the metric
\begin{equation}
ds^2 = -f^{2}(\rho) \, dt^2 + d\rho^{2} + h^{2}(\rho) \,
\left[ d\phi + a(\rho) \, dt \right]^{2} \label{m1} 
\end{equation}
and depending on whether the cosmological constant $\Lambda=-1/\ell^2 < 0$
is present or not, the metric functions are given by either \footnote{The 
constants $\beta_{0}$ and $\beta_{3}$ were set equal to 1 in \cite{ds}. Here
we keep them for later convenience.} 

\noindent 
i) nonvanishing cosmological constant:
\begin{eqnarray}
f(\rho) & = & f_{0} \, e^{2 \rho/\ell} \, X^{-1/2} \, , \quad
h(\rho) = h_{0} \, X^{1/2} \, , \quad
a(\rho) = - a_{0} + k \, \frac{f_{0}}{h_{0}} \, e^{2 \rho/\ell} \, X^{-1} \, , 
\nonumber \\
X(\rho) & \equiv & \beta_{0} + \beta_{1} \, e^{2 \rho/\ell} + 
\beta_{2} \, e^{(1/\ell - \mu k) \, \rho} \, , 
\label{coz1}
\end{eqnarray}
or

\noindent
ii) vanishing cosmological constant:
\begin{eqnarray}
f(\rho) & = & f_{0} \, Y^{-1/2} \, , \quad
h(\rho) = h_{0} \, Y^{1/2} \, , \quad
a(\rho) = - a_{0} + k \, \frac{f_{0}}{h_{0}} \, Y^{-1} \, , 
\nonumber \\
Y(\rho) & \equiv & \beta_{3} \, e^{- \mu k \rho} - \mu \, \beta_{4} \, 
(\omega_{0} + k \, \rho) \, .
\label{coz2}
\end{eqnarray}
Here $f_{0}$, $h_{0}$, $a_{0}$, $\beta_{i}$ ($i=0,1, \dots, 4$)
and $\omega_{0}$ are all real constants that arise from the integration of 
the field equations whereas $k=\pm 1$ is a free parameter that comes from the
solution of the Killing spinor equation on the supersymmetry side. 

In \cite{ds}, it was impossible to explicitly invert the functional relation
$r=h(\rho)$ for the case of the nonvanishing cosmological constant so that
the metric could be brought to the well-studied BTZ form \cite{btz}, and the
vast literature on that metric could be suitably adopted for an analysis 
of the physical meanings of the integration constants above. Instead, a 
much more complicated analysis was carried out by studying the quasilocal 
mass and the quasilocal angular momentum which was developed in \cite{do} 
in an AdS background. We refer the reader to \cite{ds} for the details of 
this.

Here a brief remark stating the differences between these quasilocal
charges and the gravitational charges in the sense of \cite{db1, db2} are
in order perhaps: The quasilocal energy in a spatially bounded region
(such as an asymptotically AdS background for our case) is defined as 
minus the `time' rate of change of the classical gravitational action.
An analogous definition exists also for the quasilocal angular momentum.
(Please see \cite{byork} and the references therein for the attempts to
define ``quasilocal gravitational charges''.) The definition of 
gravitational charges in TMG, however, are much more natural since these
gauge invariant conserved (global) charges are forged into being by
the Gauss law and the presence of asymptotic Killing symmetries 
\cite{db1, db2}.

For the time being, let us concentrate on the case of nonvanishing
cosmological constant $\Lambda = -1/\ell^2 \neq 0$. By substituting the
metric functions (\ref{coz1}) in the metric (\ref{m1}), one obtains
\begin{equation}
ds^2 = d\rho^2 - \frac{f_{0}^{2} \, e^{4\rho/\ell}}{X(\rho)} \, dt^2
+ \frac{f_{0}^{2} \, e^{4\rho/\ell}}{X(\rho)} \left(
dt + k \, \frac{h_{0}}{f_{0}} \, e^{-2\rho/\ell} \, X(\rho) \,
(d\phi - a_{0} \, dt) \right)^{2}  \label{m2}
\end{equation}
after some simplifications.

In \cite{ds}, it was found that the quasilocal mass was $a_{0}$ times the
quasilocal angular momentum (see (39) of \cite{ds}) and the asymptotic 
behavior of the metric was examined through the metric function $a(r)$
and hence $a_{0}$. It was shown that for $a$ to vanish asymptotically 
as $r \rightarrow \infty$, $a_{0}$ had to be chosen
either as 0 or as $k f_{0}/(h_{0} \beta_{1})$. Whether $a_{0}$ has a 
physical meaning or not (and whether it can be set equal to zero or not),
one should be able to make the simple change of variable 
$d\theta = d\phi - a_{0} \, dt$ in the metric (\ref{m2}). The outcome is
simply
\[ ds^2 = d\rho^{2} + 2 k \, f_{0} \, h_{0} \, e^{2\rho/\ell} \, dt \, 
d\theta + h_{0}^{2} \, X(\rho) \, d\theta^{2} \, . \]
Another simple redefinition of the coordinates as $u=k f_{0} t$ and 
$v=h_{0} \theta$ can always be made at this stage and one arrives at the
final form
\begin{equation}
ds^2 = d\rho^2 + 2 e^{2\rho/\ell} \, du \, dv + \left( \beta_{0} + 
\beta_{1} \, e^{2 \rho/\ell} + \beta_{2} \, e^{(1/\ell - \mu k) \, \rho}
\right) \, dv^{2} \, .
\label{met}
\end{equation}
It is obvious that one of the integration constants in (\ref{met}) can be
set to 1 by simple coordinate rescalings. The curvature invariants of this
metric can be calculated easily: the Ricci scalar $R=-6/\ell^{2}$ and
$R_{\mu\nu} \, R^{\mu\nu} = 12/\ell^{4}$, moreover this solution is 
asymptotically AdS (for $1/\ell - \mu k < 0$) with no curvature 
singularities. When $\beta_{0}=\beta_{2}=0$, the metric is the AdS metric 
in the Poincar\'{e} coordinates. For this case, even when one starts 
with $\beta_{1}=0$, one can still introduce it back by a simple coordinate 
redefinition as $\tilde{u}=u - \beta_{1} v/2$.

There is yet another alternative way to understand the emergence of the
constants $\beta_{0}$ and $\beta_{1}$ in the expression for $X(\rho)$.
These two terms can be thought of as describing a gravitational wave in AdS.
One can use the technique developed by Garfinkle and Vachaspati \cite{gar} 
which permits the addition of a wave to an already existing solution 
when there is a null Killing vector present. A detailed discussion of this 
method, its extension to various supergravity theories and to theories
that include nontrivial matter couplings can be found in \cite{myers}. One
can easily follow the footsteps of \cite{myers} and conclude without
any difficulties that the Garfinkle-Vachaspati method is also applicable
to the TMG theory. A brief outline of this technique can also be found
in the appendix of \cite{sad} and here we will use the notation outlined 
there. In our case, one starts from (\ref{met}) with $X(\rho)=0$, {\it i.e.} 
the AdS metric
\begin{equation}
ds^2 = d\rho^2 + 2 e^{2\rho/\ell} \, du \, dv \, \label{back1} 
\end{equation}
which satisfies $R_{\mu\nu} = (-2/\ell^2) g_{\mu\nu}$. Using the null
Killing vector $k^{\mu} = (\partial/\partial v)^{\mu}$, the scalar $\Omega$
(of \cite{sad}) is calculated easily as 
\( \Omega = \Omega_{0} \, e^{-2 \rho/\ell} \), where $\Omega_{0}$ is an 
arbitrary constant. Following relevant steps, one obtains
\( \Phi(\rho) = \Phi_{1} - \ell \, \Phi_{0} \, e^{-2 \rho/\ell} \), where
$\Phi_{0}$ and $\Phi_{1}$ are arbitrary real constants. Using these, defining
\( -\Omega_{0} \, \ell \Phi_{0} \equiv \beta_{0} \) and 
\( \Omega_{0} \, \Phi_{1} \equiv \beta_{1} \), one obtains (\ref{met}) with
$\beta_{2}$ set equal to zero in $X(\rho)$.

In fact the metric with $\beta_{2}=0$ has showed up earlier in different
contexts: It corresponds to a generalized Kaigorodov metric \cite{kai}.
It is obtainable from the AdS metric by an $SL(2,R)$ transformation 
\cite{brec} and its equivalence to the extremal limit of the BTZ black
hole \cite{btz} can be shown \cite{brec, pope}. A crucial point is that
the boundaries of the AdS and the extremal BTZ metric are different
(see \cite{pope} for details). When $\beta_{2} \neq 0$, one can again
remove the constant $\beta_{0}$ by a shift in the $\rho$ coordinate.

Another observation that needs to be stated is that in fact the constants
$\beta_{0}$, $\beta_{1}$ and $\beta_{2}$ can be taken as arbitrary
functions of $v$ and the metric (\ref{met}) is also a solution to the
TMG equations with a cosmological constant when $X(\rho)$ is replaced by
\[ X(\rho, v) \equiv \beta_{0}(v) + \beta_{1}(v) \, e^{2 \rho/\ell} + 
\beta_{2}(v) \, e^{(1/\ell - \mu k) \, \rho} \, .\]
These arbitrary functions can be thought of as describing the profile of
the gravitational wave then. However when they are left arbitrary, it is
highly probable that the supersymmetry is completely broken.

Let us now calculate the gravitational charges associated with the metric
(\ref{met}) using the procedure outlined in section \ref{conchar}. It is
clear that the metric (\ref{back1}) can be used as the background with
its timelike 
\( \bar{\xi}^{\mu} = (- \partial/\partial u + \partial/\partial v)^{\mu} \)
and spacelike
\( \bar{\zeta}^{\mu} = (\partial/\partial u + \partial/\partial v)^{\mu} \)
Killing vectors yielding the energy and the angular momentum, respectively.
After a tedious calculation, one finds that the integrand
(that is, the terms inside the parentheses) in (\ref{charge}) is given by
\[ E(\rho) = \delta^{\mu}\,_{u} \, \delta^{i}\,_{\rho} \, 
\frac{1}{2 \mu \ell^2} \, \left\{ 4 \beta_{0} (1+ \mu \ell) + \beta_{2} \,
e^{(1/\ell - \mu k) \, \rho} \, (1+ \mu k \ell) \left[ 1+ (k+2) \mu \ell
\right] \right\} \, , \]
and obviously depending on the sign of $1/\ell - \mu k$, one finds that
the energy at the boundary of AdS $(\rho \rightarrow \infty)$ is 
\footnote{Here we have chosen the Newton constant $G$ in (\ref{charge}) 
accordingly.}
\[ E = \left\{
\begin{array}{cl}
2 (k/\ell) (1+k) (\beta_{0} + \beta_{2}) \, , & 1/\ell - \mu k = 0 \\
2 \beta_{0} \, (1+ \mu \ell)/(\mu \ell^2) \, , & 1/\ell - \mu k < 0
\end{array} \right. \, . \]
As for the angular momentum, one finds that it is equal to the energy: $L=E$.

The steps that have been taken up until this point can also be repeated in
an analogous fashion for the case of vanishing cosmological constant. The
metric that corresponds to (\ref{m2}) in such a process is simply found as
\begin{equation}
ds^2 = d\rho^{2} + 2 dt \, d\theta + \left( \beta_{3} \, e^{- \mu k \rho} 
- \mu \, \beta_{4} \, (\omega_{0} + k \, \rho) \right) \, d\theta^{2} \, .
\label{fmet}
\end{equation}
If one starts from the flat metric \( ds^2 = d\rho^{2} + 2 dt \, d\theta \)
and applies the Garfinkle-Vachaspati method to add a gravitational wave to
this spacetime, one readily finds $\Omega = \Omega_{0}$ and 
$\Phi(\rho) = \Phi_{1} + \Phi_{0} \, \rho$, where $\Phi_{0}$, $\Phi_{1}$ 
and $\Omega_{0}$ are arbitrary real constants, and these, with the
definitions $\Omega_{0} \, \Phi_{1} \equiv - \mu \beta_{4} \, \omega_{0}$
and $\Omega_{0} \, \Phi_{0} \equiv - \mu k \, \beta_{4}$, lead to the metric
(\ref{fmet}) with $\beta_{3}$ set equal to zero in $Y(\rho)$.

Once again the constants $\beta_{4}$, $\omega_{0}$ and $\beta_{3}$ can be
taken as arbitrary functions of $\theta$ and the metric (\ref{fmet}) is 
also a solution to the TMG equations when $Y(\rho)$ is replaced by
\[ Y(\rho, \theta) \equiv \beta_{3}(\theta) \, e^{- \mu k \rho} 
- \mu \, \beta_{4}(\theta) \, (\omega_{0}(\theta) + k \, \rho) \, . \]
These arbitrary functions can be thought of as describing the wave profile
again, but with these functions in place, it may be that there is no 
supersymmetry left to preserve then.

As for the gravitational charges related to (\ref{fmet}), the background
to work with is simply the flat metric 
\( ds^2 = d\rho^{2} + 2 dt \, d\theta \) and the Killing vectors needed 
are just the timelike
\( \bar{\xi}^{\mu} = (-\partial/\partial t + \partial/\partial \theta)^{\mu} \)
and the spacelike
\( \bar{\zeta}^{\mu} =(\partial/\partial t + \partial/\partial \theta)^{\mu} \)
vectors. The steps leading to (\ref{charge}) can easily be repeated by 
setting $\Lambda=0$ in the relevant places and replacing the covariant
derivatives with respect to $\bar{g}_{\mu\nu}$ with ordinary derivatives.
One then finds that the terms inside the parentheses in (\ref{charge})
is given by 
\[ E(\rho) = \delta^{\mu}\,_{t} \, \delta^{i}\,_{\rho} \, \left(
\mu k \, \beta_{4} + \frac{1}{2} \, \mu \, \beta_{3} \, (2k-1) \,
e^{-\mu k \rho} \right) \, , \]
and depending on the sign of $\mu k$, one finds the energy as 
$\rho \rightarrow \infty$ to be
\[ E = \left\{
\begin{array}{cl}
0 \, , & \mu = 0 \\
\mu k \beta_{4} \, , & \mu k > 0
\end{array} \right. \, . \]
The angular momentum is again given by $L=E$.

\section{\label{conc} Conclusions}

In this paper, we showed how the physical properties of a known 
supersymmetric solution of the full cosmological TMG theory can 
be better understood by the Garfinkle-Vachaspati method and then 
determined its conserved charges for bulk asymptotically flat and 
constant curvature backgrounds. Even though the question of how 
the supersymmetric version of the TMG theory can be obtained from 
any compactification of M-theory or, for that matter, any higher 
dimensional supergravity theory remains open, provided that an
exact form of the CFT dual of TMG can be formulated on the boundary
of AdS, this particular supersymmetric solution should be suitable
for understanding the AdS/CFT duality in the infinite momentum frame
\cite{brec}, \cite{pope}. Another open question that deserves attention 
is the problem of finding a supersymmetric matter coupled extension 
of the TMG theory. Looking for the charged versions of the metrics
studied here within this model and their relations with the ones 
presented in \cite{ds2} would be certainly worth the effort.

\section{\label{ackno} Acknowledgments}

We thank G. Cl\'{e}ment for useful discussions. This work is partially 
supported by the Scientific and Technical Research Council of Turkey 
(T\"{U}B\.{I}TAK); the work of B.T. is also supported by the ``Young 
Investigator Fellowship'' of the Turkish Academy of Sciences (T\"{U}BA) 
and by a T\"{U}B\.{I}TAK Kariyer Grant.


\begin{thebibliography}{99}
\bibitem{des} S. Deser, R. Jackiw and S. Templeton, Phys. Rev. Lett. {\bf 48}
975 (1982); Ann. Phys. (N.Y.) {\bf 140} 372 (1982); {\bf 185} 406(E) (1988).
\bibitem{dba1} S. Deser and B. Tekin, Class. Quantum Grav. {\bf 19} L97 
(2002). 
\bibitem{deskay} S. Deser and J.H. Kay, Phys. Lett. {\bf 120B} 97 (1983); 
S. Deser, in {\it Quantum Theory of Gravity: Essays in Honor of the 60th 
Birthday of Bryce S. DeWitt}, edited by S.M. Christensen (Adam Hilger, London,
1984), pp. 374-381.
\bibitem{btz} M. Ba\~{n}ados, C. Teitelboim and J. Zanelli, Phys. Rev. Lett.
{\bf 69} 1849 (1992);  M. Ba\~{n}ados, M. Henneaux, C. Teitelboim and 
J. Zanelli, Phys. Rev. D {\bf 48} 1506 (1993).
\bibitem{any} S. Deser, Phys. Rev. Lett. {\bf 64} 611 (1990).
\bibitem{vuo} I. Vuorio, Phys. Lett. {\bf 163B} 91 (1985).
\bibitem{per} R. Percacci, P. Sodano and I. Vuorio, Ann. Phys. (N.Y.) 
{\bf 176} 344 (1987).
\bibitem{ort} M.E. Ortiz, Class. Quantum Grav. {\bf 7} L9 (1990).
\bibitem{par} A. Edery and M.B. Paranjape, Phys. Lett. {\bf 413B} 35 (1997).
\bibitem{nub} Y. Nutku and P. Baekler, Ann. Phys. (N.Y.) {\bf 195} 16 (1989).
\bibitem{nut} Y. Nutku, Class. Quantum Grav. {\bf 10} 2657 (1993).
\bibitem{gur} M. G{\"u}rses, Class. Quantum Grav. {\bf 11} 2585 (1994).
\bibitem{cle} G. Cl\'{e}ment, Class. Quantum Grav. {\bf 11} L115 (1994).
\bibitem{ds} T. Dereli and \"{O}. Sar{\i}o\u{g}lu, Phys. Rev. D {\bf 64}
027501 (2001).
\bibitem{dal} K. Ait Moussa, G. Cl\'{e}ment and C. Leygnac, Class. 
Quantum Grav. {\bf 20} L277 (2003).
\bibitem{db1} S. Deser and B. Tekin, Class. Quantum Grav. {\bf 20} L259 
(2003).
\bibitem{db2} S. Deser and B. Tekin, Phys. Rev. Lett. {\bf 89} 101101 (2002);
Phys. Rev. D {\bf 67} 084009 (2003).
\bibitem{dkt} S. Deser, \.{I}. Kan{\i}k and B. Tekin, Class. Quantum Grav. 
{\bf 22} 3383 (2005).
\bibitem{mtz} C. Mart\'{i}nez, C. Teitelboim and J. Zanelli, Phys. Rev. D 
{\bf 61} 104013 (2000). 
\bibitem{do} T. Dereli and Yu. N. Obukhov,  Phys. Rev. D {\bf 62} 024013 
(2000).
\bibitem{byork} J.D. Brown and J.W. York, Jr, Phys. Rev. D {\bf 47} 1407 
(1993).
\bibitem{gar} D. Garfinkle and T. Vachaspati, Phys. Rev. D {\bf 42} 1960 
(1990); D. Garfinkle, Phys. Rev. D {\bf 46} 4286 (1992). 
\bibitem{myers} N. Kaloper, R.C. Myers and H. Roussel, Phys. Rev. D {\bf 55}
7625 (1997).
\bibitem{sad} N.S. De\u{g}er and \"{O}. Sar{\i}o\u{g}lu, J. High Energy Phys.
{\bf 12} 039 (2004).
\bibitem{kai} V.R. Kaigorodov, Dokl. Akad. Nauk. SSSR {\bf 146} 793 (1962)
[Sov. Phys. Doklady {\bf 7} 893 (1963)].
\bibitem{brec} D. Brecher, A. Chamblin and H.S. Reall, Nucl. Phys. B {\bf 607}
155 (2001).
\bibitem{pope} M. Cveti\v{c}, H. Lu and C.N. Pope, Nucl. Phys. B {\bf 545} 309
(1999).
\bibitem{ds2} T. Dereli and \"{O}. Sar{\i}o\u{g}lu, Phys. Lett. {\bf 492B} 339
(2000).
\end{thebibliography}
\end{document}